\documentstyle[prl,aps,preprint]{revtex}
\begin{document}
\draft
\title{Recovery of the persistent current induced by the 
electron-electron interaction in mesoscopic metallic rings}
\author{G. Chiappe}
\address{ Departamento de F{\'\i}sica, Universidad de Buenos Aires, 
Ciudad Universitaria, \\
Pabell\'on {I}, 1428 Capital Federal, Buenos Aires, Argentina}
\author{J.A. Verg\'es \cite{now} and E. Louis}
\address{ Departamento de F{\'\i}sica Aplicada, Universidad de Alicante,\\
Apartado 99, E-03080 Alicante, Spain.}
\date{Received 9 February 1996; accepted xx XXX 1996 by F. Yndurain}
\maketitle
\begin{abstract}
Persistent currents in mesoscopic metallic rings induced by static
magnetic fields are investigated by means of a Hamiltonian
which incorporates diagonal disorder and the electron-electron interaction
through a Hubbard term ($U$). Correlations are included up to second order
perturbation theory which is shown to work accurately for $U$ of
the order of the hopping integral. If disorder is not very strong,
interactions increase the current up to near its value
for a clean metal. Averaging over ring lengths eliminates the first
Fourier component of the current and reduces its value, which
remains low after interactions are included.

\end{abstract}

\vspace{2 cm}
{\bf Keywords:} A. nanostructures D. electronic transport 
D. electron-electron interactions

\newpage
The theoretical prediction concerning the existence of persistent
currents in isolated rings in the presence of static
magnetic fields \cite{1} has recently been experimentally demonstrated
\cite{2}. As a consequence of the boundary conditions imposed by the
magnetic field, both the current and the free energy are periodic
functions of the magnetic flux threading the loop $\Phi$ with a
fundamental period $\Phi_0 = h/e$. In clean (ideal) metallic loops at
zero temperature, this current is expected to be nearly $ev_F/L$,
where $L$ is the perimeter of the loop. Impurities are thought to
decrease this current in an amount $l/L$, where $l$ is the mean
free path. A persistent current periodic in the magnetic flux 
with period $\Phi_0$ was reported in \cite{2}. However, the value
of this current resulted to be $(0.3-2.0)ev_F/L$, that is, with no
apparent correction due to disorder (impurities) and two or
three orders of magnitude higher than predicted by theoretical
analyses. On the other hand, magnetization measurements on
an ensemble of $10^7$ Cu loops \cite{3} suggested a persistent current of
about $3\times10^{-3}ev_F/L$, periodic in the magnetic flux with a
period $\Phi_0/2$. The latter result is believed to be a consequence
of the large number of rings \cite{4,45}: averaging eliminates the first
Fourier component of the current although the second survives.

The large current found in a single isolated ring has prompted several
theoretical studies aiming to find a mechanism which could compensate
the effect of impurities \cite{45,5,6,7,71,72,73,74,75}. In particular 
it has been suggested that
e-e interactions may tend to homogenize the system,
offsetting the reduction in the current promoted by disorder. 
Calculations by means of spinless fermions models indicate that 
e-e interactions reduce the value of the persistent 
current \cite{5,6,7}, or show increased currents in unrealistic 
regions of the parameter space \cite{6}. On the other hand, recent analysis
of the effect of a Hubbard term ($U$) indicate that the interaction
compensates in part the detrimental effect of
disorder \cite{71,72,73,74}. In particular renormalization group and
exact calculations \cite{71,72} show that the Drude stiffness 
of the disordered system increases with $U$. This result is in line
with calculations of the full current by means of a first order
perturbation treatment of the Anderson-Hubbard Hamiltonian \cite{73,74}; 
we note, however, that first order perturbation theory 
may strongly overestimate the effects of the interactions.   


In this work we present the results of a study of persistent 
currents in one dimensional metallic rings 
by means of the Anderson-Hubbard Hamiltonian. 
Correlation is included up
to second order perturbation theory \cite{8}, which is shown to work
very accurately for the rather small values of $U$ 
which apply for a metallic band. The Hamiltonian used in this work is:

\begin{eqnarray}
H = \sum_{l\sigma} \epsilon_{l} c_{l\sigma}^{\dagger}c_{l\sigma}
-t\sum_{<lm>\sigma}\left[  e^{2\pi i \Phi /L} 
c_{l\sigma}^{\dagger} c_{m\sigma} + H.c. \right]  
\nonumber
\end{eqnarray}
\begin{eqnarray}
+ U \sum_{l}n_{l\uparrow}n_{l\downarrow} ~~~~~,
\end{eqnarray}

\noindent
where the on-site energies, $\epsilon_{l\sigma}$, are chosen 
randomly between $-W/2$ and $W/2$.
$U (>0)$ is the on-site Coulomb repulsion, $t$ the hopping
integral between nearest neighbor sites, and $L$ the number of sites.
The results herewith discussed correspond to one electron per site
(half-filling) which is adequate to describe metallic systems.
In the following we will take $t=1$ and all lengths
will be given in units of the lattice constant. Second
order perturbation theory is implemented as follows. 
Single particle excitations
are included in all orders of perturbation theory (Hartree selfconsistency)
\cite{81} whereas two particle excitations will be included in second order
(Fig. 1). Only paramagnetic solutions 
are considered. It should be noted that
this approach could likely be valid to describe the Anderson phase
of this system (small $U$) \cite{9}. Actually, a transition to a
Hubbard-like phase is expected for large $U$ \cite{9}
which cannot be described within
the second order perturbation theory utilized in this work. Note, however,
that in the present systems $U$ is always smaller than half the bandwidth.


The persistent current will be calculated from:

\begin{equation}
I(\Phi) = -\frac{1}{2\pi}\frac{\partial E_g(\Phi)}{\partial \Phi} ~~~~~,
\end{equation}

\noindent
where $E_g(\Phi)$ is the ground state energy for a flux $\Phi$.

To facilitate the choice of $W$, we 
have calculated the localization length from the decay 
of the Drude weight with the system size at $U=0$. The Drude weight $D$
is given by:

\begin{equation}
D = \frac{L}{4\pi^2}\left( \frac{\partial^2 E_g(\Phi)}{\partial \Phi^2}
\right)_{\Phi=\Phi_m} ~~~~~~,
\end{equation}

\noindent
where $\Phi_m$ is the location of the minimum of the ground
state energy. We have carried out simulations for $L=4-512$, 
keeping constant the product $L\times$number of realizations (16384). 
The results can be very accurately
fitted by exponential functions $D(L) \propto \exp(-(L/\xi)$ for the three 
$W$ shown in Fig. 2. The resulting values for the localization length,
$\xi$, are 110, 29 and 7 for $W$=1, 2 and 4, respectively. The accuracy
of these results is supported by the fact that similar scaling laws
were obtained for other magnitudes. For instance $<I^2>^{1/2}$ 
averaged over the magnetic flux, is very accurately fitted by
$<I^2>^{1/2} \propto [\exp(-(L/\xi)]/L$, with $\xi =28.4$ for $W$=2,
in excellent agreement with the previous result. Our numerical 
results for $\xi$ agree very well with the well-known expression
obtained at the mid-band for small disorder \cite{10}, 
$\xi = 105/W^2$ ($W << 2\pi$);
in fact this formula gives 105, 26 and 6.6, for $W$ = 1, 2 and 4,
respectively, to be comapred with the results given above. We
note that for
rings of $L \approx 100$, $W$=1,2 will represent two cases in which 
the localization length is of the order or significantly smaller 
than the length $L$ of the ring. 


The accuracy of second order perturbation theory was checked by 
comparing results for $I(\Phi)$ obtained by means of perturbation theory 
with those of exact calculations (Lanczos method) for rings 
of $L=8$ and several values of $W$ and $U$ (Fig. 3).
We note that for $U$ up to 1.5 the agreement is excellent. 
Instead the agreement for larger $U$ is poor and the numerical 
results obtained by means of perturbation theory are not
shown in the Figure. In particular, the second order results show
abrupt changes in the current (near $\Phi$ = 0 and/or 0.5) 
which we associate to changes in the nature of the wavefunction 
(Anderson to Hubbard-like); these abrupt changes in the current 
were used as an indication
of the failure of second order perturbation theory when larger rings
were investigated. The results of Fig. 3 do also illustrate
a key point of the effects of $U$. As the exact results show, for small
values of the e-e interaction an increase
in the current is observed (the effect of $U$ increases with
$W$, for instance a factor of 7 increase in $I$ is obtained 
when an interaction of $U=2$ is switched in a chain of $L$=8 
and $W$=8). Further increases in $U$ reduce the current
below its value for $U=0$, and for $U=\infty$ the
current vanishes no matter the degree of disorder. 
These results can be understood in terms of
clear physical grounds \cite{11}. For small $U$ the
e-e interaction homogenizes the disordered system
promoting a delocalization of the electronic states and, subsequently,
an increase of the current. When $U$ is increased above a given value,
the scattering mechanism added by the e-e interaction
dominates and the current starts to decrease, becoming zero for an
interaction of infinite strength. 
Previous analyses of these systems did not consider this
question in detail \cite{12}. 

Figure 4 shows the results for rings of length $L$=100. In order to
facilitate the comparison with the experimental data for a 
single metallic loop \cite{2}, the results of a single realization are 
shown. The results for the ordered ring and $U$=0 are also shown as a 
reference. We first note that for $U$=0 the current for $W$=1 is much larger
than for $W$=2. This is a
consequence of the fact that $\xi/L$ is 1.1 for $W$=1 and 
0.29 for $W$=2. Electron-electron interactions 
increase the magnitude of the persistent current
very significantly. Although this increase is relatively more important
for $W$=2 (up to a factor of 7 for the represented results), the current 
is still well below the value corresponding to the
ordered system. Instead for $W$=1, the current for $U$=1.5 gets rather
close to 
that of the ordered system, in qualitative agreement with the
experimental results \cite{2}. It should be pointed out that we have never
obtained currents higher than that of the ordered system, as apparently
observed in the experiments on single rings \cite{2}. In our opinion, 
e-e interactions can never increase the current beyond that of the ordered
ring, as these interactions are by themselves a source of scattering
(see above). 

One of the most intricate issues concerning persistent currents
in mesoscopic rings is the way in which averages are performed. This is
of particular significance when plausible explanations for
experiments such as the one performed on $10^7$ metallic rings
\cite{3} have to be found. Here, in order to facilitate the
investigation of the effects 


of the e-e interaction, 
we have adopted the simple approach of averaging
over the rings length. This kind of average has not been considered 
previously, despite of the fact that the actual lengths of the rings
investigated in Ref. \cite{3} should show considerable dispersion.
Fig. 5 reports the persistent currents obtained by averaging those for
rings of lengths $L$=100 and 102 (five realizations each). We first note
that the persistent current is a periodic function of the magnetic flux
on the scale of half a flux quanta ($\Phi_0$), as already predicted by
other authors. The reason for this halving of the period is, in this
case, very simple; in fact the two rings, being half-filled, have an
even or odd number of electrons per spin, and, thus, their resulting
currents are out of phase in $\Phi_0/2$. On the other hand, the
amplitude of the current is significantly smaller than in the case of a
single ring. Although as in the case of a single ring the
e-e interaction increases the current, this remains well
below the value corresponding to the ordered system. More extensive
averaging may reduce further the amplitude of the current, albeit,
as discussed in \cite{4}, it should remain finite.


In conclusion, we have presented an investigation of persistent currents
in mesoscopic metallic rings by means of a model Hamiltonian which
includes diagonal disorder and a Hubbard term. We have proved that
second order perturbation theory works very accurately for the low
values of the strength of the e-e interaction ($U$)
which are adequate in the present systems. Our results show that
for small $U$ the current is increased up to near the value corresponding
to the non-interacting case, in line with the common believe which
assumes that the interaction should homogenize the disordered
system. Instead, for large values of $U$, the e-e
interaction acts as a scattering mechanism that reduces the
current, which decays to zero in the $U = \infty$ limit. These
results may explain the large currents observed in isolated metallic 
rings. We have also shown that averaging over ring length 
promotes a halving of the period of the current
and a strong reduction of its amplitude (not substantially increased
by the e-e interaction), in line with the experimental
information obtained on $10^7$ Cu loops.

\acknowledgments

We wish to acknowledge financial support from  the spanish CICYT (grant
MAT94-0058). G. Chiappe wishes to thank the "Generalitat Valenciana"
for financial support.

\newpage

{\large\centerline{FIGURE CAPTIONS}}

Fig.1: Feynman diagram included in the second order perturbation
theory utilized in this work.

Fig.2: Scaling of the Drude weight ($D$) with the length of the ring ($L$) in the
non-interacting case ($U$=0) and for several values of the disorder parameter.
$W$=0.01 (circles), 1 (squares), 2 (up triangles) and 4 (down triangles).

Fig.3: Persistent currents in rings of length $L=8$, half-filling
and $W$=4 as a function of the magnetic flux threading the ring.  
$U$= 0 (thin continuous line). Exact results (thick lines): $U$= 1, 1.5,
3 and 4, dotted, continuous, dashed, and chain 
lines, respectively. Second order perturbation theory results: $U$= 1 
(circles) and 1.5 (triangles).

Fig.4: Persistent current in rings of $L$=100 and half-filling as a
function of the magnetic flux threading the ring.
a) $W$=1 and $U$=0 and 1 (continuous and dashed thick 
lines, respectively). The results for the ordered chain 
(thin continuous line) are also shown (for $U$=0). b) $W$=2 and 
$U$=0, 1 and 2 (continuous, dashed and dot-dashed 
thick lines, respectively). The e-e interaction was treated 
in second order perturbation theory (see text).

Fig.5: Persistent current in metallic rings as a function of the
magnetic flux threading the ring. The results correspond to the average
of those obtained for two rings of $L$=100 and 102 (5 realizations each),
both at half filling. Results for $W$=1 and $U$=0 and 1 
(continuous and dashed lines, respectively) are shown.


\begin{references}
\bibitem{81}
Note that this approximation is already beyond the first order 
perturbation theory used in Refs. \cite{73,74}.
\bibitem{12}
Note that although the qualitative behavior of the current versus $U$
was correctly described in \cite{6}, the actual increments in its
amplitude were much lower than those obtained here, likely due to
the spinless fermions model and/or the small rings investigated in
\cite{6}. On the other hand, the maximum in 
the current versus $U$ is not accounted for by means of renormalization
group equations which do not incorporate $4k_f$ impurity scattering
\cite{71}, as discussed in \cite{72}. We finally note that this
maximum was already observed in the behavior of the localization
length for the Anderson-Hubbard Hamiltonian \cite{11}.
\end{references}
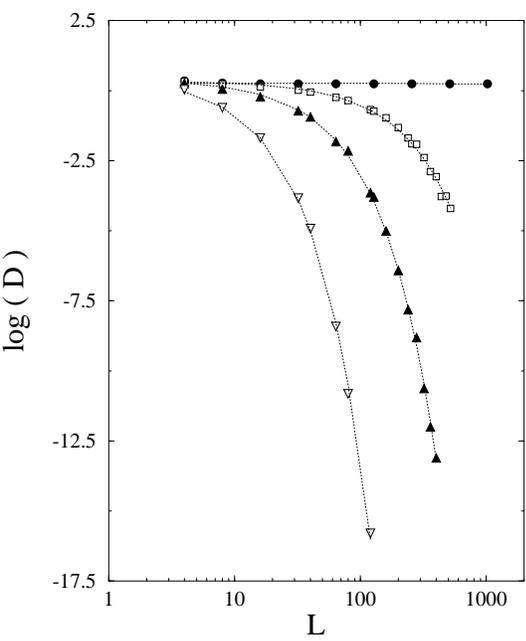
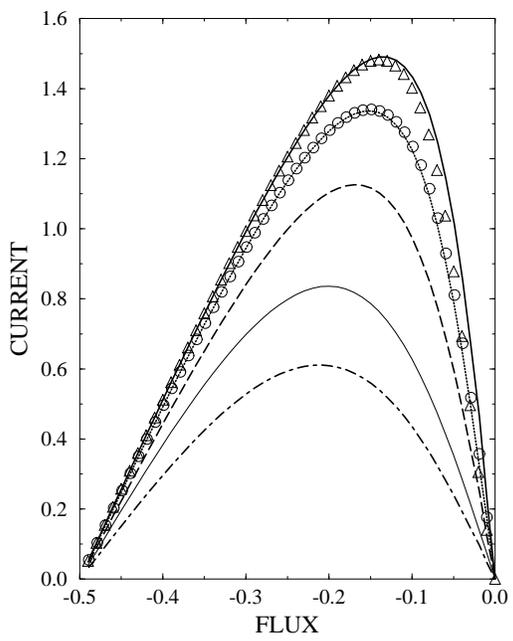
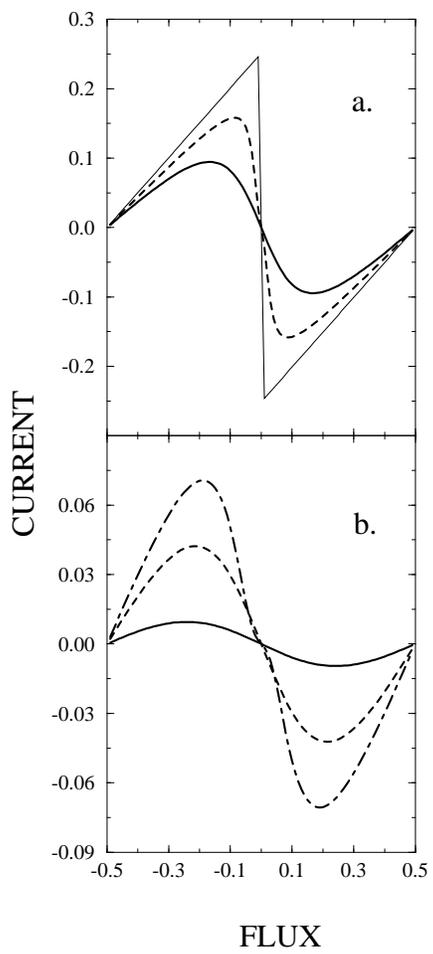
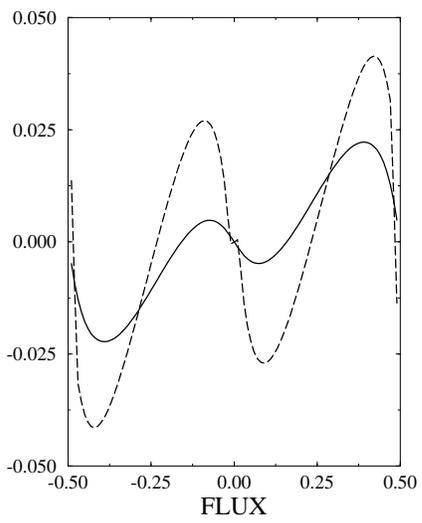
\end{document}